\documentclass[prl,twocolumn,showpacs,reprint,preprintnumbers,superscriptaddress,tightenlines]{revtex4-1}                                         
\usepackage[T1]{fontenc}  
\usepackage{amsfonts} 
\usepackage{amsmath,amsbsy,amssymb,graphicx}                   
\usepackage{times} 
\usepackage{color}

\begin{document}
\title{Relaxation to Negative Temperatures in Double Domain Systems}                
\author{Yusuke~Hama} 
\affiliation{RIKEN Center for Emergent Matter Science (CEMS), Wako, Saitama 351-0198, Japan} 
\affiliation{National Institute of Informatics, 2-1-2 Hitotsubashi, Chiyoda-ku, Tokyo 101-8430, Japan}
\author{William~J.~Munro}  
\affiliation{NTT Basic Research Laboratories, NTT Corporation, 3-1 Morinosato-Wakamiya, Atsugi, Kanagawa, 243-0198, Japan }        
\affiliation{National Institute of Informatics, 2-1-2 Hitotsubashi, Chiyoda-ku, Tokyo 101-8430, Japan}
\author{Kae~Nemoto}
\affiliation{National Institute of Informatics, 2-1-2 Hitotsubashi, Chiyoda-ku, Tokyo 101-8430, Japan}
\date{\today} 
\begin{abstract}{ 
We investigate the relaxation of two collective spins in double domain system, which are individually coupled to a single bosonic reservoir,  by varying the total number of spins in each domain and their initial spin configurations. A particularly interesting situation occurs when the spin domains are set in an antiparallel configuration. Further for an unbalanced configuration where the number of spins are in the excited state initially  is much greater than that in the ground state, the spin ensemble prepared in the ground state relaxes towards a negative-temperature state.    }
\end{abstract}

\maketitle

{\it Introduction}.---
In recent years, the hybridization of quantum systems has become a key technique to design and demonstrate novel quantum behaviors \cite{hybrid1,hybrid2,circuitqedreview1}.  With the rapid progress in quantum coherent manipulation, hybrid quantum systems have now entered the regime where we can observe unexpected or rather counterintuitive behavior even in the presence of imperfections and noise \cite{pretheremalization,matsuzaki,strongcoupling1,strongcoupling2}. 
Such hybrid systems have not only shown the capability to achieve superior properties each individual system alone cannot achieve \cite{qubus,matsuzaki}, 
but they also shred light on the fundamental complexity of such quantum systems including coupling structures and decoherence mechanisms \cite{jared}.   Currently a number of such hybrid systems have been proposed (and realized in some cases) with various elements coming from atomic molecular \& optical systems to solid-state systems. Example include for instance, trapped ions \cite{trappedionsreview1,trappedionsreview1}, optical cavities \& resonators \cite{cavityqedreview1,cavityqedreview2}, 
electron and nuclear spin ensembles in quantum dots (QD) or nitrogen-vacancy centers in diamond \cite{quantumdotreview1,nvcenterreview1,nvcentermechresonatorreview1},
 superconducting circuits in quantum electrodynamic systems \cite{circuitqedreview1} and mechanical  resonators \cite{nvcentermechresonatorreview1,mechanicalresonatorreview1}. 
This large diversity of component systems really allows one to explore the unique space hybridization potentially allows.
 
Our focus in this letter will be on collective coherent behavior in  such hybrid spin systems.  Coherent collective coupling was originally proposed in quantum optics with one notable example being the superradiant decay of a spin ensemble collectively coupled with  an optical mode \cite{superradiance1,GH82}. More recently squeezing of a spin ensemble via collective decoherence has been proposed \cite{shane}. 
When those spins couple collectively with a bosonic reservoir, the spin dynamics dramatically changes from those driven by individual spin-boson coupling.  
Typically investigations of such collective behavior have focused on single spin ensembles, primarily as they are theoretically more tractable and easier to experimentally realize.
Here we examine a hybrid system of a spin ensemble and a bosonic reservoir beyond this regime. This is achieved by introducing spin domains in the ensemble. Spin domains allows one to partition the ensemble into distinct non interacting components that yet still couple to the same bosonic reservoir.   Hybrid systems with multiple domains are being realized experimentally,  for instance  a nuclear spin ensemble in a semiconductor materials can couple with the Nambu-Goldstone bosonic mode in the quantum Hall regime \cite{Kumadaetal,Fauziprb,yhamaetal}.

\begin{figure}[b] 
\includegraphics[width=0.37\textwidth]{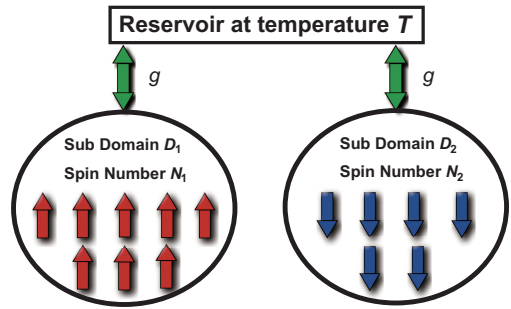}
\caption{ Schematic representation of a double spin domain system coupled to a single reservoir.  Here we denote the  first spin domain containing $N_1$ spins as $D_1$ where the spins are shown as red arrows, the second spin domain $2$ with $N_2$ spins is labelled as $D_2$ with the spins represented by the blue arrows. In both domains, each spin couples with the bosonic reservoir (at temperature $T$) with the coupling constant $g$.  } 
\label{spindomainsystemfnl} 
\end{figure}  

Our illustrative hybrid system is schematically depicted in Fig. \ref{spindomainsystemfnl} as an ensemble of spin-1/2 particles coupled to a single bosonic reservoir. The ensemble is decomposed into two independent domains, each characterized by its initial state. In our investigation here, we focus on the dynamics of the double domain system, however want to note that the analysis can be easily expanded to multi-domain systems.   We explore the dynamics of these double domain system to show that collective quantum phenomena, such as superradiant decay in a domain as well as in the total system.  Further we show that the relaxation of a domain to a negative temperature is possible.
  
 \begin{figure}[b]
\includegraphics[width=0.48\textwidth]{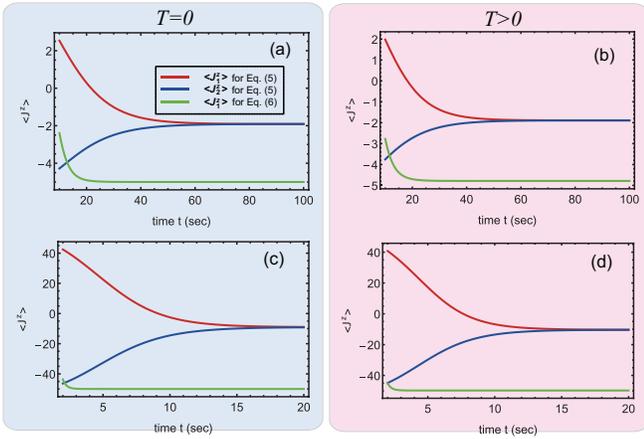}
\caption{ The collective spin relaxations for balanced number configurations. The red curves are the relaxations for $J^z_1$ while blue curves are for $J^z_2$. 
The dotted green curves are the relaxations under the initial state where both the spins in $D_1$ and $D_2$ are in the excited states. (a) Zero-temperature result for $N_1=N_2=10.$  (b) Finite-temperature $T=400$ mK results for $N_1=N_2=10.$  (c) Zero-temperature result for $N_1=N_2=100.$  (d) Finite-temperature result for $N_1=N_2=100$ with $T=400$ mK.}
\label{balancefnl} 
\end{figure}

{\it Modelling}.--- 
Let us now turn our attention to a mathematical description of this hybrid system and our modeling of it. As mentioned earlier, our total ensemble (as shown in Fig. \ref{spindomainsystemfnl}) is divided into  two spin domains labeled as domain $D_1$ and $D_2$. Each domain $D_{1(2)}$ contains $N_{1(2)}$ 1/2-spins with the same frequency $\omega_{\text{s}}$.  In each domain all spins are initially aligned in the same direction either  upwards or downwards along the $Z$-axis. The spins are not in the symmetric state of the whole system (double domain)  but are for each individual domain. Thus we can regard the domain $D_{1(2)}$ as collective spin $J_{1(2)}$ whose spin magnitude is $N_{1(2)}/2$. Next, the spins in domains $D_1$ and $D_2$ couple with the bosonic reservoir with a coupling constant $g$. We further assume that the spin domain size is much smaller than the wavelength to ensure the uniformity of the coupling in terms of both amplitude and phase. We can now describe our overall system (two domains and the reservoir) by the Hamiltonian
\begin{align}
H&  =
\hbar\omega_{\text{s}} (J^z_1+J^z_2)+\int d^d k \;E_{\boldsymbol{k}}r^\dagger_{\boldsymbol{k}}r_{\boldsymbol{k}}\notag\\
&+\frac{\hbar g}{2} \left[(J^+_1+J^+_2)R+(J^-_1+J^-_2)R^\dagger\right], 
\label{hamiltonian1} 
\end{align}
where the first term represents the free energy of the two domains ($a=1,2$ represents the first and second domain respectively) with $J^{x,y,z}_a$ being the usual $x,y,z$ collective spin operators,
$d$ is the spatial dimension of the system, {and $\boldsymbol{k}=(k_1,\ldots,k_d)$ the wave vector of the reservoir.
The raising (lowering) operators of these collective spins are defined by  $J^{\pm}_a=J^x_a\pm iJ^y_a$. Next the second Hamiltonian term represents the free energy of the reservoir where $E_{\boldsymbol{k}}$ is the linear dispersion relation of the reservoir with $r _{\boldsymbol{k}}(r^\dagger_{\boldsymbol{k}})$ it's annihilation (creation) operator satisfying the commutation relation $[r _{\boldsymbol{k}},r^\dagger_{\boldsymbol{k}^\prime}]=\delta(\boldsymbol{k}-\boldsymbol{k}^\prime)$. The final Hamiltonian terms represents the coupling of the strength $g$ between the two domains and the reservoir  where $R=\int  d^d k \kappa_{\boldsymbol{k}} r_{\boldsymbol{k}}$ with $\kappa_{\boldsymbol{k}}$ being a continuous function of $\boldsymbol{k}$ whose exact form depends on the system under consideration.

In our modeling, we will set $\omega_{\text{s}}/2\pi=10$ GHz and  take $g\ll \omega_{\text{s}}$ so that second order perturbation with respect to the interaction Hamiltonian between the spins and reservoir is valid. Using a perturbation approach, we can derive the master equation for the reduced density matrix  composed only of the two domains defined by $\rho_{\text{S}}(t)=$Tr$_R(W(t))$ with $W(t)$
the density matrix for the total system and Tr$_R$ representing the operation tracing out the reservoir degrees of freedom. Further we will assume that initially the spin domains and the reservoir are uncorrelated. We will characterize the reservoir by the density matrix $\rho_{\text{R}}=\exp(-\beta H_{\text{R}})/(\int d^d k \exp(-\beta H_{\text{R}}))$ where $H_{\text{R}}$ is the second term in Eq.   \eqref{hamiltonian1}, $\beta=1/k_{\text{B}}T$ with
$k_{\text{B}}$  being Boltzmann constant and $T$ the reservoir temperature. Now from \eqref{hamiltonian1}, the master equation using the Born-Markov approximation can be written as \cite{carmichaeltxb}
\begin{align}
\dot \rho_{\text{S}}(t)&=
-i\omega_{\text{s}} 
\left[ J^z_1+J^z_2 ,  \rho_{\text{S}}(t)\right]+\gamma(\bar{n}+1){\cal L}(J^-_1+J^-_2)\notag\\
&\;\;\;\;\;\;\;+ \gamma\bar{n}{\cal L}(J^+_1+J^+_2),
\label{smasterequation1}
\end{align}
where ${\cal L}(A)= 2A \rho A^\dagger - A^\dagger A \rho - \rho A^\dagger A $, $\bar{n}=1/(e^{\beta\hbar\omega_s}-1)$ is the Bose-Einstein distribution and $\gamma$ is the damping rate which is a function of both the coupling $g$ and $|\kappa_{\boldsymbol{k}}|^2$ at the wave vector  $\boldsymbol{k}_s$ (satisfying $E_{\boldsymbol{k}_s}=\hbar\omega_{\text{s}}$). Here we take $\gamma=0.01$ Hz. Now from our master equation \eqref{smasterequation1} the equations of motion for the expectation values of  collective spins are

\begin{align}
\frac{d}{dt}\langle
J^{z}_{1(2)} \rangle&=-2 \gamma(2\bar{n}+1) \langle J^{z}_{1(2)} \rangle \notag\\
&+\frac{\gamma}{2}\left(-N_{1(2)}(N_{1(2)}+2)+4\langle J^{z}_{1(2)} \rangle^2-2\langle A_{12} \rangle \right), \notag\\
\frac{d}{dt}\langle A_{12} \rangle &=-2 \gamma(2\bar{n}+1)\left( \langle A_{12} \rangle-4 \langle J^{z}_{1}J^{z}_{2}\rangle\right) \notag\\
&+2\gamma\left(\langle J^{z}_{1 } \rangle+\langle J^{z}_{2}\rangle \right)\left(\langle A_{12} \rangle-2\langle J^{z}_{1}J^{z}_{2}\rangle\right) \notag\\
&+\gamma\left( N_{2 }(N_{2 }+2) \langle  J^{z}_{1} \rangle  +N_{1 }(N_{1 }+2) \langle  J^{z}_{2 } \rangle \right)\notag\\
\frac{d}{dt}\langle J^{z}_{1}J^{z}_{2} \rangle &= -\frac{1}{2}\frac{d}{dt}\langle A_{12} \rangle ,\label{doublespinevequations1} 
\end{align}
where $A_{12}=J^+_{1}J^-_{2}+J^-_{1}J^+_{2}$. Further in order to derive Eq.  \eqref{doublespinevequations1}, we have used the approximation \cite{approximation}
\begin{align}
&\langle (J^z_{i})^2\rangle\approx\langle 
J^z_{i}\rangle^2, \;\;\;\; \langle J^z_i(J^z_j)^2\rangle\approx \langle J^z_i J^z_j  \rangle   \langle J^z_j\rangle,\notag\\
 &\langle J^z_{i} J^{\pm}_{i}J^{\mp}_{j}\rangle\approx  \langle  J^{\pm}_{i}J^{\mp}_{j}\rangle\langle J^z_{i}\rangle\pm\langle  J^{\pm}_{i}J^{\mp}_{j}\rangle,\label{approximation1} \\
&\langle  J^{\pm}_{i}J^z_{i}J^{\mp}_{j}\rangle\approx  \langle  J^{\pm}_{i}J^{\mp}_{j}\rangle\langle J^z_{i}\rangle ,\notag
\end{align}
with $i,j=1,2$ ($i\neq j$). The collective spin relaxations in the double spin domain system are described by four dynamical variables:  $\langle J^z_{1}  \rangle, \langle J^z_{2}  \rangle, \langle A_{12}  \rangle,$  and $\langle J^z_1 J^z_2  \rangle.$  $\langle A_{12}  \rangle$ describes the spin flip-flop between the two domains and its dynamics may be identified with that of the correlation between the $J^z_{1}$  and $J^z_{2}$ as seen in \eqref{doublespinevequations1}. Let us now analyze the relaxation processes under the various conditions in terms of the initial spin-domain configuration and domain numbers $N_1$ and $N_2$.

{\it Collective Spin Relaxations}.---
We present in Figs. \ref{balancefnl}-\ref{zero104102fnl} our results for the relaxation processes for different conditions. Here we mainly focus on such relaxations under an antiparallel configuration initial state 
\begin{align}
|\text{SDAP}\rangle &=|\uparrow\ldots\uparrow\rangle_{D_1}\otimes|\downarrow\ldots\downarrow\rangle_{D_2}. 
\label{initialstate}
\end{align}
To understand the essence of these spin relaxation processes clearly, we focus on the steady state behavior and the relaxation times for each domain. In the case of the balanced system, that is, domains of equal spin size, the results shown in Fig. \ref{balancefnl} indicate that starting with the antiparallel configuration (\ref{initialstate}) the relaxation process for each domain are similar, decaying to the steady state with the same $\langle J^z_{i}\rangle_{i=1,2}$. The average number of excitations at the steady state is dependent on both the domain size $N_{i}$ and of course the reservoir temperature.  Here, the red curves are the relaxation processes for the first domain $D_1$ while blue curves are those for second domain $D_2$. As a comparison, the green curve shows the decay process under the parallel configuration initial state defined by 

\begin{align}
|\text{SDP}\rangle &=|\uparrow\ldots\uparrow\rangle_{D_1}\otimes|\uparrow\ldots\uparrow\rangle_{D_2}.
\label{initialstate2}
\end{align}
As both this initial state (\ref{initialstate2}) and the Hamiltonian (\ref{hamiltonian1}) satisfy the symmetry of SU(2) for the total collective spin, the total state decays on the symmetric subspace.  In this case, the collective effect of the decay, i.e. superradiant, is most prominent.  This superradiant effect would be more visible when the spin size gets larger, which is numerically shown in the difference between (a) for $N_1=N_2=10$ and (c) for $N_1=N_2=100$ in  Fig. \ref{balancefnl}  (The relaxation in a finite temperature as shown in (b) and (d) exhibit a similar behavior.)  
To determine the dependency of the relaxation time $\tau_N$ for the antiparalell configuration \eqref{initialstate} on the domain size, we plot $\tau_N$ in Fig. \ref{sizedependency}  versus $N=N_1=N_2$ for both zero (a) and a finite (b) temperature. $N$ was varied between 100 - 1000. Curve fitting to these data points shows $\tau_N$ has the form $a/N+b$, a clear signature of superradiant decay \cite{GH82}.

\begin{figure}[htb]
\includegraphics[width=0.48\textwidth]{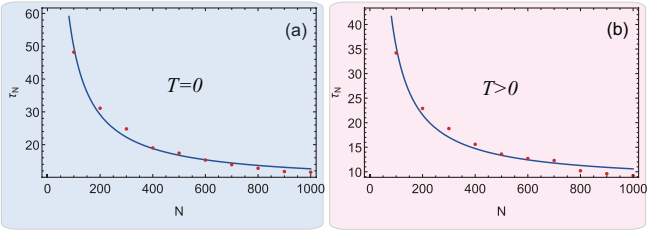}
\caption{Plot of the relaxation time $\tau_N$ for the initial configuration  \eqref{initialstate} as a function of domain size $N$ for zero (a) and finite (b) temperature. Curve fitting indicates a functional form for $\tau_N \sim a/N+b$ with $a=4125.85,b=8.51$ for (a) $a=2756.58,b=7.84$ for (b). The 1/$N$ dependence is a clear signature of superradiant decay. } 
\label{sizedependency} 
\end{figure}
The steady state of the antiparalell configuration for each domain contains more excitations than the parallel configuration case.  
This is due to the smaller spin size for the each domain ($N_{1,2}<N_1+N_2$) and the violation of the SU(2) symmetry at the initial time of the dynamics.  We can always write the antiparallel configuration as a sum of a symmetric subspace component and an non-symmetric subspace component. The symmetric subspace component decays like in the parallel configuration case, however the non-symmetric subspace component decays differently. This will be illustrated in more detail next when we discuss the unbalanced case $N_1\neq N_2$.

\begin{figure}[htb] 
\includegraphics[width=0.47 \textwidth]{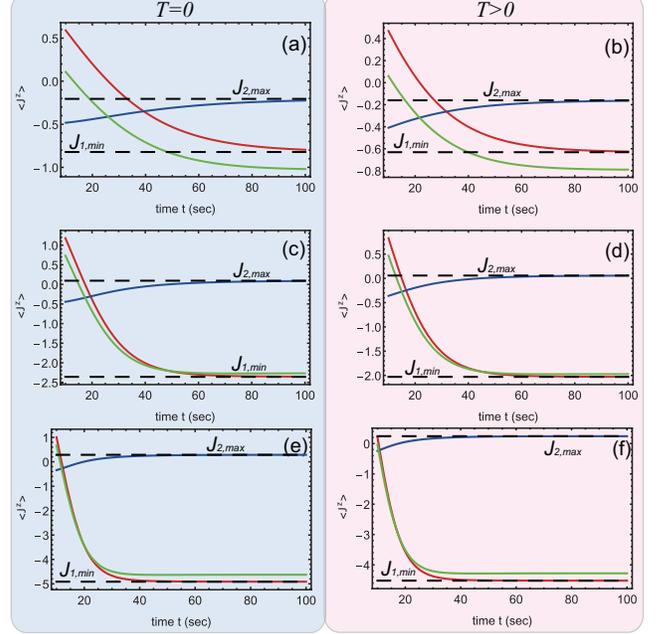}   
\caption{ Plot of the collective spin relaxations for various unbalanced number configurations ($N_1>N_2$) at zero and finite (400mK) temperature. The configuration considered are $N_1=2, N_2=1$ at zero (a) and finite (b) temperature, $N_1=5, N_2=1$ at zero (c) and finite (d) temperature, and $N_1=10, N_2=1$ at zero (e) and finite (f) temperature. The red curves, blue curves, and green curves  represents $\langle J^z_1\rangle$, $\langle J^z_2\rangle$, and $\langle J^z_1\rangle+\langle J^z_2\rangle$, respectively. $J^z_{1,min}$ ($J^z_{2,max}$) are the steady-state solution for $J^z_{1(2)}$.  }
\label{unbalancefnl} 
\end{figure}

In Fig. \ref{unbalancefnl} we plot the expectation values $\langle J^z_1\rangle$ (red), $\langle J^z_2\rangle$ (blue), and $\langle J^z_1\rangle+\langle J^z_2\rangle$ (green) for various unbalanced configuration $N_1>N_2$ with $N_1=2,5,10$ and $N_2=1$ at both zero and finite (400mK) temperature.  In these unbalanced spin size cases, the smallest domain $D_2$ shows completely different relaxation compared to the larger domain $D_1$. In fact, although $D_2$ was initialized in the ground state of the domain as given by \eqref{initialstate}, this subsystem relaxed into more highly excited states (even at zero temperature).  
More interestingly, (c) and (d) for $N_1=5, N_2=1$ show that the second domain decays to a steady state at an effective negative temperature (the population is greater than 50\%).  In fact, the steady state of the second domain gradually gets closer to an almost fully excited state when the system size gets larger.  This tendency can be seen from (c) and (e) (or (d) and (f)).  
To illustrate this point further, we plot in Fig. \ref{zero104102fnl} the relaxation process for $N_1=10^4$ with $N_2=10^2$ at zero-temperature.  It can be clearly seen that in the steady state the second domain is approaching the fully excited state, despite superradiant relaxation of the first domain leading it to its ground state with $J^z_1 \sim-N_1/2$.  

\begin{figure}[htb]
\includegraphics[width=0.47\textwidth]{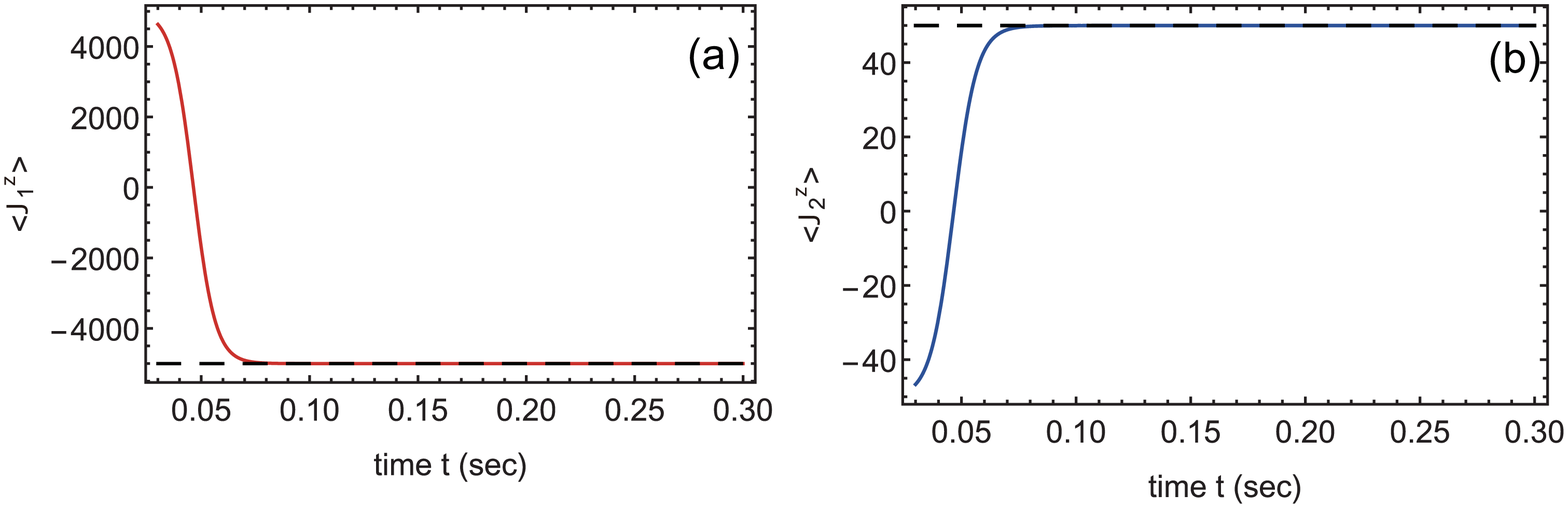}
\caption{Plot of the collective spin relaxations $\langle J^z_1\rangle$ (red), $\langle J^z_2\rangle$ for the unbalanced configuration at zero-temperature with (a) $N_1=10^4$ and (b) $N_2=10^2$.  }
\label{zero104102fnl}  
\end{figure}
The relaxation to a negative temperature state occurs even when the second domain is initially in its ground state with no excitation.  It is interesting that this can be seen even in a system as small as several spins, though this may appear rather counter intuitive. However we need to remember several things: first and foremost is that we have an energy exchange coupling to the reservoir and so at zero temperature the decay to a negative temperature is not due to the energy flow for the first domain to the second domain via the reservoir. Second this is not a direct effect from the subradient states of the total system.  The initial states we considered are not subradient states, which do not decay with the coupling to the reservoir, and if a subradient state was prepared, the system should be effectively decoupled to the reservoir. Therefore this behavior must be a result of the collective decay of the total spin system.

Let us explore this decay characteristics of the collective coupling even with a small system considered of $N_1=5, N_2=1$.  With an initial state $|\uparrow_1\ldots\uparrow_{5}\rangle_{D_1}\otimes|\downarrow_6\rangle_{D_2}$, we need to consider two manifolds of the total system: that is, the symmetric subspace of the spin size $(N_1+N_2)/2$ and the second subspace of the spin size $(N_1+N_2-2)/2$.  The dynamics conserves the spin size, these two subspaces are enough to represent the system through the entire dynamics, and there is no transfer of the population between these subspaces.  The initial component in each manifold decays via the coupling to the reservoir, hence the component on the symmetric manifold will decay to the ground state $|\downarrow_1\ldots\downarrow_{6}\rangle$. Similarly the component on the second manifold will decay to its ground state. The decay process is confined in each subspace and hence gives the steady states the excitations in the second domain.  This mechanism of collective decay also may generate entanglement between the two domains at the steady state, even though the initial state is separable.

{\it Discussion and Conclusion}.---  
In this letter, we have investigated collective spin relaxation processes in a double spin-domain system where all spins in the two domains are collectively coupled to a bosonic reservoir such as the Nambu-Goldstone (NG) mode in electron spin systems.   The dynamics of the total system of course shows superradiant decay.  Although the initial states we choose were not optimal to observe the supreradiant decay, they were sufficient to demonstrate its nature and more importantly are easier to experimentally realize. When the two domains are not balanced and the unbalance is large enough, we may see the relaxation to a negative temperature. This decay behavior appears more prominent when the size difference becomes large.  
 
The model we discussed in this letter should be able to be implemented in various physical systems, but importantly the decay to  negative temperatures needs to be observed in time scales shorter than the spin dephasing time.  One possible realization is with a nuclear spin ensemble in GaAs semiconductors coupled with the NG mode as a low-frequency electron spin fluctuation \cite{Kumadaetal,Fauziprb}.  Here the spin dynamics in quantum dots and the quantum Hall regime has been extensively investigated \cite{quantumdotreview1,electronnuclear1,electronnuclear2,electronnuclear3}.   
 It is known that the total system of a nuclear spin ensemble and the NG mode can be described by the Dicke model \eqref{hamiltonian1} \cite{yhamaetal}.  
 The nuclear spins can be polarized initially by the dynamic nuclear polarization \cite{electronnuclear1,electronnuclear2,electronnuclear3,DNP1,DNP2,DNP3}, but may not be fully polarized in one direction, which would lead to more than a single domain coupled to the NG mode.  

Finally our analysis also indicates that one should be careful when ignoring systems coupled to reservoir.  Usually we assume  ancillary systems coupled only to the reservoir can be ignored, however as our analysis showed, when there are other systems equally coupled to the same reservoir, they can affect the system of interest significantly and fundamentally.  
Even the temperature of the shared reservoir is fixed, one needs to be careful to define temperature for the system at hand, as we have seen, the temperature of each domain appears to be different.  These effects are much more prominent with the shared reservoir in comparison to the dynamics seen in the systems coupled with independent reservoirs.  The case of shared reservoir, the physical properties of the steady state, such as temperature and entanglement, is highly dependent on the initial state of the total system, which might include ancillary systems hidden in the reservoir coupled with it in the same way.
In particular, when a small domain is imbedded into  a larger domain, a local hot spot might appear as the result of the spin relaxation.  

\acknowledgements
We thank Mohammad Hamzah Fauzi, Yoshiro Hirayama and Emi Yukawa for fruitful discussions and comments.  This work was supported in part by the RIKEN Special Postdoctoral 
Researcher Program (Y.~H),  the JSPS KAKENHI Grant Number 25220601 and the MEXT KAKENHI Grant number 15H05870 (K.~N).

\end{document}